\documentclass[usenatbib]{mn2e_modmargin}
\usepackage{amsmath}
\usepackage{multirow}
\usepackage{url}
\usepackage{subfigure}
\usepackage{color}
\usepackage{graphicx}
\usepackage{subfig}
\usepackage{txfonts}
\usepackage{natbib}

\title{Planet-disk interaction in highly inclined systems}
\author[Hanno Rein]{Hanno Rein\thanks{E-mail: \texttt{rein@ias.edu}}\\
Institute for Advanced Study, 1 Einstein Drive, Princeton, NJ 08540, USA}
\begin{document}

\date{Received: 10 June 2011; Accepted: 3 March 2012  }

\pagerange{\pageref{firstpage}--\pageref{lastpage}} \pubyear{2012}

\maketitle

\label{firstpage}

\begin{abstract}
We study the interaction of a proto-planetary disk and a planet on a highly inclined orbit in the linear regime. 
The evolution of the planet is dominated by dynamical friction for planet masses above several Earth-masses. 
Smaller planets are dominated by aerodynamic drag, especially for very high inclinations and retrograde orbits. 

The time-scales associated with migration and inclination damping are calculated.
For certain values of the inclination, the inclination damping time-scale is longer than the migration time-scale and the disk lifetime.
This result shows that highly inclined planets can not (re-)align with the proto-planetary disk. 

We discuss the dependence of numerical simulations on the gravitational softening parameter. 
We find only a logarithmic dependence, making global three dimensional simulations of this process computationally feasible.

A large fraction of Hot Jupiters is on highly inclined orbits with respect to the rotation axis of the star. 
On the other hand small-mass planetary systems discovered by the Kepler mission have low mutual inclinations.
This shows that there are two distinct formation mechanisms at work.
The process that creates inclined Hot Jupiters does not operate on small mass planets because the damping timescales are so long that these systems would still be inclined today.
\end{abstract}


\section{Introduction}
The number of confirmed extra-solar planets was at 556 when this paper was submitted\footnote{By the time this paper was accepted the number has gone up to 759.}.  
Many of them, mostly the high-mass, close-in Hot Jupiters, are on rather extreme orbits, especially when compared to the solar system. 
Using the Rossiter-McLaughlin effect, several planets have been found to be in highly inclined orbits with respect to the sky-projected spin axis of the star \citep[][and references therin]{Triaud2010,Simpson2011}. 
Although still debated \citep{Lai2011}, it is generally assumed that the proto-planetary disk is roughly aligned with the spin axis of the host star. 

Several mechanisms have been proposed to account for planets on orbits with high inclinations \citep[e.g.][]{FabryckyTremaine2007,Chatterjee2008}. 
Here, we do not address the question on how the planet got on the orbit in the first place.

Before highly inclined planets had been found, one of the main areas of theoretical work was the interaction of planets and proto-planetary disks.
The robust result is planetary migration, although the precise speed and also the direction are still being debated and most likely depend strongly on the global disk structure \citep{Paardekooper2010a,Paardekooper2011b}.
This naturally leads to the question of how a planet on a highly inclined orbit interacts with a proto-planetary disk.
Several authors have looked at the evolution of moderately inclined planets in three dimensional simulations \citep{Marzari2009,Bitsch2011,Cresswell2007}

In this paper, we consider a planet that is on a highly inclined orbit, not embedded in the disk, and crossing a vertically stratified disk twice per orbit. 
The planet creates a density perturbation in the disk. 
These perturbations interact gravitationally with the planet. 
This interaction is known as dynamical friction.
Furthermore, aerodynamical drag from accreting material might become important.

In section \ref{sec:linear} we first calculate the linear response of the disk and the resulting friction force on the planet analytically. 
We then estimate the relevant time-scales and compare the effect to aerodynamical drag. 
In section \ref{sec:numerical} we describe numerical simulations verifying the analytic result. 
We finally discuss the consequences in section \ref{sec:conclusions}.

\section{Linear Theory}\label{sec:linear}

\begin{figure}
\centering
\resizebox{\columnwidth}{!}{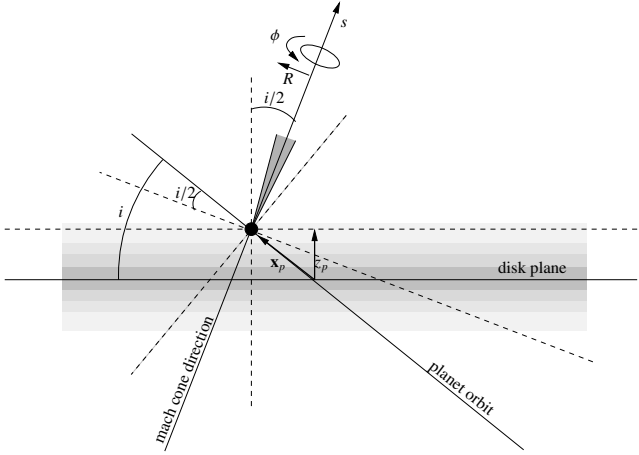}
\caption{Illustration of the setup; a planet with inclination $i$ passing through the disk on a highly inclined orbit from top left to right bottom.\label{fig:setup}}
\end{figure}

We assume that the disk is in Keplerian rotation with angular velocity $\Omega$ at a distance $a$ from the host star with mass $M_\star$. 
We further assume that the planet is on a circular orbit with velocity $v_k=\Omega a$ and inclination $i$. 
The relative velocity between the disk and the planet is then $v_{imp}=2v_k\sin(i/2)$. 
The force due to dynamical friction or aerodynamic drag felt by the planet is anti-parallel to the velocity of the planet relative to the velocity of the disk. This setup is illustrated in Fig.~\ref{fig:setup}. The force has an angle $i/2$ to both the normal direction of the planet's orbit and the normal direction of the disk. We ignore the shear and curvature of the disk, and assume that there is no net force in the radial direction. Without loss of generality, we will work in units where $v_k=1, a=1, GM_\star=1$ unless otherwise stated.

\subsection{Dynamical friction}
When the planet comes close to the disk plane, it induces density perturbations in the disk. 
These then exert a gravitational force on the planet. 
The force is directed so that the relative velocity of the disk and the planet always decreases. 
Thus the name, dynamical friction.  

\subsubsection{Linearized equation}
In analogy to \cite{Ostriker1999} we start by defining the perturbed density~$\alpha$ and velocity~$\beta$ by~$\rho(\mathbf{x},t)\equiv\rho_0(\mathbf{x})[1+\alpha(\mathbf{x},t)]$ and~$\mathbf{v}(\mathbf{x},t)\equiv c_s\beta(\mathbf{x},t)$. 
We consider an isothermal equation of state with constant sound-speed $c_s$. 
The linearized continuity and momentum equations in a frame moving with the disk are given by
\begin{align}
\frac{\partial \alpha}{\partial t} + c_s \nabla \mathbf{\beta} &= 0\\
\frac{\partial \mathbf{\beta}}{\partial t} + c_s \nabla \alpha + c_s^{-1} \nabla \Phi  &=0, 
\end{align}
where $\Phi=-Gm/|\mathbf{x} - \mathbf{x}_p|$ is the potential due to the planet at position~$\mathbf{x}_p$.
The linear equations are valid in the limit that~$\alpha,|\mathbf{\beta}|,|\nabla \rho_0/\rho_0|\ll1$. 
As shown by \cite{KimKim2009}, this is justified as long as the dimensionless nonlinearity parameter $\eta$ is less than unity. 
For an Earth-like planet on a $90^\circ$ orbit we have
\begin{align}
\eta \equiv \frac{Gm}{c_s^2\,r_p\,(M^2-1)} \sim 0.03.
\end{align}
The background steady state is a stratified disk in the~$z$~direction with $\rho_0(\mathbf{x})=\bar\rho_0\,\exp[-z^2/H^2]$, where~$H$ is the thickness of the disk. 
In a Keplerian accretion disk we have $H=\sqrt{2}\,c_s\Omega^{-1}$. 
By eliminating~$\beta$, the above equations lead to 
\begin{align}
\nabla^2 \alpha -\frac1{c_s^2} \frac{\partial^2 \alpha}{\partial t^2} = -\frac1{c_s^2}\nabla^2\Phi.
\end{align}
This can be solved with a Green's function \citep{Ostriker1999}. The steady state solution is given by 
\begin{align*}
\alpha(\mathbf{x},t) &= \frac{Gm/c_s^2}{\sqrt{s^2+R^2(1-M^2)}} \cdot 
	\left\{ \begin{array}{ll}
	1 & M<1\\
	2 & M>1 \text{ and } s/R < - \sqrt{M^2-1}\\
	0 & \text{otherwise.}
	\end{array} \right.\label{eq:alphaost}
\end{align*}
The above equation describes a Mach cone in the supersonic case with no perturbations outside of the cone (see Fig.~\ref{fig:setup}). 
$s$ is the coordinate in the direction of the Mach cone and $R$ is in the direction perpundicular to it. 
Because these equations are linear, we can easily generalize this result to a softened planet potential of the form $\Phi(R,s)=-Gm/\sqrt{R^2+s^2+b^2}$ with softening length $b$. 
For this potential, the corresponding mass distribution is given by
\begin{align}
\rho_p(R,s) &=  \frac{3\,m\,b^2}{4\pi\,\left(b^2+R^2+s^2\right)^{5/2}}.
\end{align}
Thus, the density perturbation created by this mass distribution is
\begin{align}
\bar{\alpha}(\mathbf{x},t) & = \iiint \alpha(\mathbf{x-x'},t) \;\frac{\rho_p(|\mathbf{x'}|)}m \, d^3\mathbf{x'}. \label{eq:alphabar}
\end{align}
As we will integrate over $\mathbf{x}$ in the following section, we can simply use Fubini's theorem and do not need to evaluate the integral here.

\subsubsection{Change in orbital energy after one passage through the disk}
The specific gravitational force exerted on the planet from an unperturbed, infinitely large stratified sheet with density profile $\rho_o(\mathbf{x})$ is symmetric to the mid-plane. 
Because of this symmetry, there is no net change in orbital energy of the planet after one crossing.

However, the force from the perturbed density $\alpha$ is not symmetric, the planet therefore changes its specific orbital energy after each passage trough the disk.
This can be calculated as an integral over the specific force $\mathbf{F}$ along the path of the planet $\mathbf{x}_p=(x_p,y_p,z_p)^{-1}$,
\begin{align*}
\Delta E 
=& \int \mathbf{F}(\mathbf{x}_p)\cdot d\mathbf{x}_p\\
=& \int \mathbf{F}(\mathbf{x}_p(t))\cdot \left(\begin{array}{c}0\\\-\cos(i)\\\sin(i)\end{array}\right) \;dt .
\end{align*}
Using the definitions from the last section (see also Fig.~\ref{fig:setup}), we can change the coordinate system and describe any point $\mathbf{x}=(x,y,z)^{-1}$ with $R, s$~and~$\phi$ using the relations
\begin{align*}
x &= x_p + R \sin(\phi)\\
y &= y_p + R \cos(\phi)\cos(i/2)+s \sin(i/2)\\
z &= z_p + R \cos(\phi)\sin(i/2)+s \cos(i/2).
\end{align*}
The force on the planet is then given by
\begin{align*}
\mathbf{F}(\mathbf{x}_p) &=- G \iiint \frac{\rho(\mathbf{x})\; (\mathbf{x}_p-\mathbf{x})}{\left|(\mathbf{x_p}-\mathbf{x})^2+b^2\right|^{3/2}}d^3\mathbf{x} \\
&=- \iiint \frac{2G^2m\bar\rho_0/c_s^2}{\sqrt{s^2+R^2(1-M^2)}} e^{-z^2/H^2}\frac{\mathbf{x}_p-\mathbf{x}}{\left|s^2+R^2+b^2\right|^{3/2}} R\;dR\;ds\;d\phi.
\end{align*}
The change in orbital energy can then be integrated over $\phi$ and the trajectory of the planet, such that
\begin{align*}
\Delta E 
=& \int \mathbf{F}(\mathbf{x}_p)\cdot d\mathbf{x}_p \\
=& -\iiiint \frac{2G^2m\bar\rho_0c_s^{-2}e^{-z_p^2/H^2}}{\sqrt{s^2+R^2(1-M^2)}} \frac{\mathbf{x}_p-\mathbf{x}}{\left|s^2+R^2+b^2\right|^{3/2}} R\;dR\;ds\;d\phi\;d\mathbf{x}_p\\
=& -4\pi^{3/2} H G^2m\bar\rho_0/c_s^2\frac{\sin(i/2)}{\sin(i)} \quad \Lambda(M,r_>,b)/M^2,
\end{align*}
where we have defined the remaining two integrals as 
\begin{align*}
\Lambda(M,r_>,b) =&  \iint \frac{M^2\;s\;R}{\left[s^2+R^2(1-M^2)\right]^{1/2}\left|s^2+R^2+b^2\right|^{3/2}} \;dR\;ds. 
\end{align*}
The integral has the form of a Coulomb logarithm and we have introduced an additional cut-off scale at large distances (e.g. the disk size), $r_>$.
This allows us to solve the Coulomb integral exactly,
\begin{align}
\Lambda(M,r_>,b)&= -\sqrt{M^2-1} \cdot\text{ atan}\left(\frac{r_>}{\sqrt{(M^2-1) (b^2+r_>^2)}}\right)\nonumber\\&\quad-\log\left(\frac{\sqrt{b^2+r_>^2}-r_>}{b}\right). \label{eq:lambdalong}
\end{align}
Note that if we further introduce the surface density $\Sigma = H \sqrt{\pi} \bar\rho_0$ and substitute the expression for the Mach number, we can rewrite the specific change in orbital energy as
\begin{align}
\Delta \hat E_\alpha &=-\frac{\pi G^2m \Sigma }{v_k^2 \sin(i/2)\sin(i)} \;\Lambda(M,r_>,b). \label{eq:deltae}
\end{align}
The result is almost independent of $H$, $c_s$ and $M$. The only dependencies are hidden in the function $\Lambda$ which behaves like a Coulomb logarithm.
$\Lambda$ is typically of order $\sim1-10$ and reflects the uncertainty about the small scale processes near the planet's surface and the large scale effects at scales where the disk curvature and shear become important.
Equation \ref{eq:deltae} reduces to that of dynamical friction in a collision-less medium in the limit of $r_>\gg b$ and $M \gg 1$. 
To see this note that we can take that limit in Eq.~(\ref{eq:lambdalong}) to get 
\begin{align}
\Lambda(M,r_>,b) &\rightarrow \log\left(\frac{r_>}{\sqrt{2}\;b}\right), \label{eq:lambdashort}
\end{align}
which should be compared to Eq.~(8.5) in \cite{BinneyTremaine2008}.

It is also worth noting that our result is qualitatively similar to those summarized by \cite{Artymowicz1994} (see his Eq.~69) but includes explicitly the dependence on the gravitational softening parameter $b$. Thus, it can be used as a benchmark in numerical simulations that make use of such a softening. Furthermore, as we will show later, we can argue that a softening parameter that is only slightly smaller than the disk scale height can be used in numerical simulations with a moderate error.

\subsubsection{Time-scales}
The rates of change in semi-major axis and inclination can be expressed in terms of the specific normal $\hat N$ and tangential force $\hat T$ \citep{Burns1976}. For a circular orbit, these are
\begin{align*}
\frac{da}{dt} = 2 \sqrt{\frac{a}{GM_\star}} \;a\;\hat T\quad\quad\text{   and   }\quad\quad
\frac{di}{dt} = \sqrt{\frac{a}{GM_\star}} \;\hat N.
\end{align*}
Averaging over one orbit, using Eq.~(\ref{eq:deltae}) and noting that the force from dynamical friction is inclined by an angle $i/2$ with respect to the normal of the orbit, we get
\begin{align*}
\Delta a &=   \frac{4a^2}{GM_\star} \; \Delta \hat E_\alpha   &&=  \frac{4\pi m \Sigma a^3}{M_\star^2} \frac1{\sin(i/2)\sin(i)}   \;\Lambda(M,r_>,b),\\
\Delta i &=   \frac{2a}{GM_\star} \;\Delta \hat E_\alpha  \tan^{-1}(i/2) &&=  \frac{4\pi m \Sigma a^2}{M_\star^2}   \frac{1}{4\sin^3(i/2)} \;\Lambda(M,r_>,b).
\end{align*}
The time-scales for inclination and semi-major axis damping are then
\begin{align}
\tau_a &= \frac{M_\star^2}{2 m \Sigma a^2 }  \sin(i/2)\sin(i)\;\Lambda^{-1}(M,r_>,b)\; \Omega^{-1} , \label{eq:taua}\\
\tau_i &= \frac{M_\star^2}{ m \Sigma a^2 }   i \sin^3(i/2)\;\Lambda^{-1}(M,r_>,b)\;\Omega^{-1}. \label{eq:taui}
\end{align}

\subsection{Aerodynamic drag}
The aerodynamic drag due to mass accretion might also be important. The accreted mass after one crossing is 
\begin{align}
\Delta m & = \frac{\pi\Sigma r_p^2 }{\cos(i/2)} .
\end{align}
The accretion is dominated by the geometrical cross section as long as the Bondi radius $r_a={2Gm}/v_{imp}^2$ is small compared to the physical radius of the planet $r_p$. For a 100 Earth mass planet with a size of 10 Earth radii on an orbit with $i=90^\circ$ we have $r_a\sim0.7r_p$. Thus the geometrical term is dominant. We will therefore assume this to be the only term for the remainder of this paper. 
The direction of the resulting force is the same as for the dynamical friction, i.e. $45^\circ$ in the $z\phi$ plane for a planet on a $90^\circ$ orbit, and the magnitude of the specific force is
\begin{align}
f_d & = \frac{\Omega \,\Delta m}{\pi\, m} v_{imp}. 
\end{align}
The relevant time-scales are then 
\begin{align}
\tau_a &= \frac{a}{\dot a} = \frac{1}{2T}\sqrt{\frac{GM_\star}{a}}\\
\tau_i &= \frac{i}{\dot i} = \frac{i}{N}\sqrt{\frac{GM_\star}{a}},
\end{align}
where the tangential and normal components of the force can be decomposed as in the previous section.
Putting everything together, we have
\begin{align}
\tau_a &= \frac{ m }{\Omega \, \Sigma\,r_p^2}  \frac{\cos(i/2)}{2\tan(i/2)} \label{eq:tauagas}\\
\tau_i &= \frac{ m }{\Omega \, \Sigma\,r_p^2}  i\cos(i/2).\label{eq:tauigas}
\end{align}
Note that the dependencies on mass and inclination are different compared to the dynamical friction time-scales.

\subsection{Comparison of time-scales}
\begin{figure}
\centering
\subfigure[10 Earth mass planet.]{\includegraphics[width=\columnwidth]{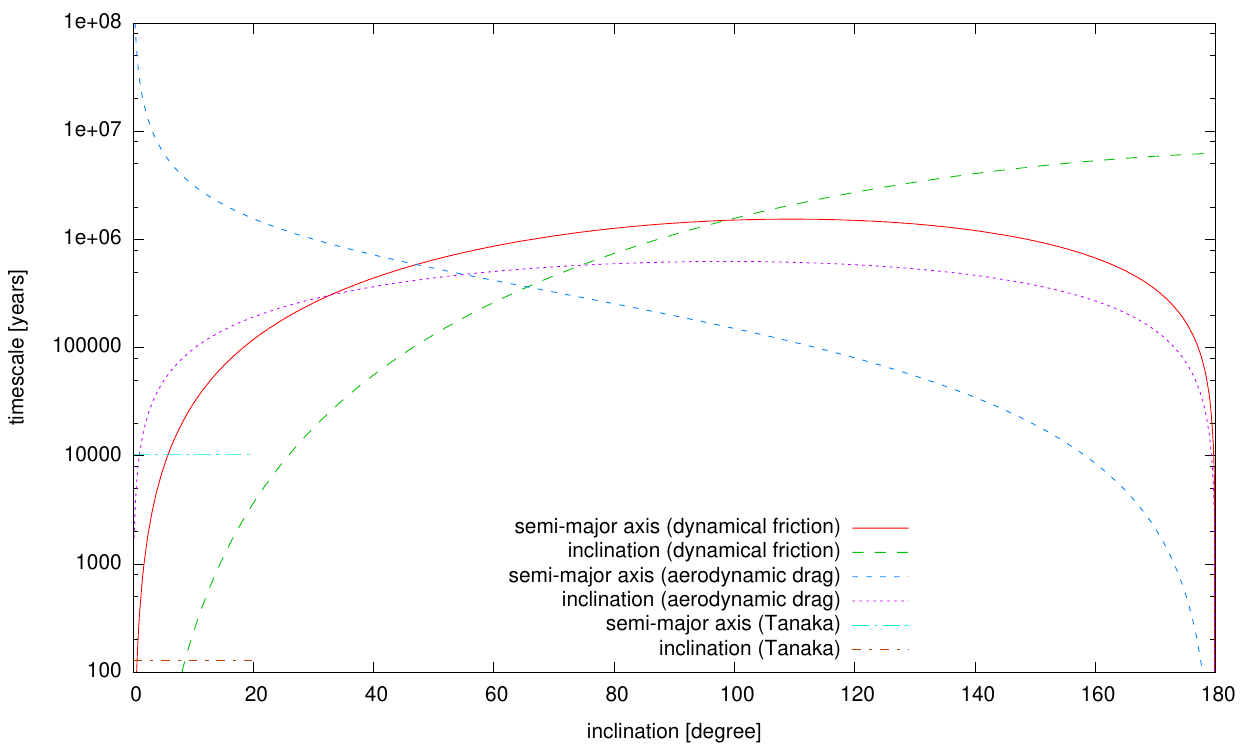} }
\subfigure[100 Earth mass planet.]{\includegraphics[width=\columnwidth]{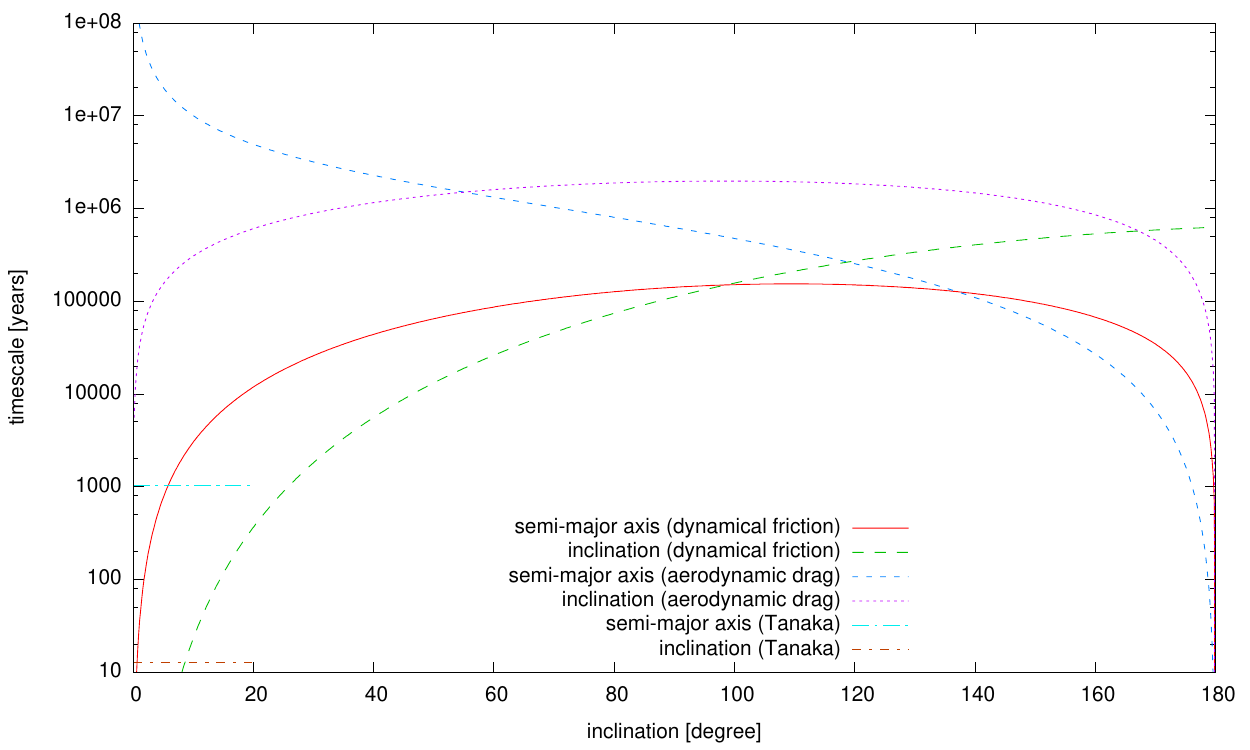}}
\caption{Migration and inclination damping time-scales.\label{fig:tau}}
\end{figure}

As an illustration, let us assume a solar nebula with a surface density $\Sigma=4200\,\text{g/cm}^2$ \citep{Weidenschilling1977} and a 10 Earth-mass planet at $1\,\text{AU}$. 
We will also set $\Lambda=4$ for simplicity. 
The time-scales given by Eqs.~(\ref{eq:taua}) and (\ref{eq:taui}) as well as the time-scales for the aerodynamic drag, Eqs.~(\ref{eq:tauagas}) and (\ref{eq:tauigas}), are plotted in the first panel of Fig.~\ref{fig:tau} as a function of the inclination $i$. We also plot the time-scales for type I migration calculated by \cite{Tanakaetal2002} and \cite{Tanakaetal2004} as a comparison. 
Their inclination damping time-scale is only valid for small mass planets embedded in the disk ($i\lesssim$~few degrees). 
\cite{Bitsch2011} showed that this is roughly consistent with numerical simulations when planets are completely embedded in the disk.

In the lower panel of Fig.~\ref{fig:tau}, we plot the same quantities for a 100 Earth mass planet. 
Here, the aerodynamic drag is reduced by a factor $\sqrt{10}$, whereas dynamical friction is increased by a factor $10$ compared to the 10 Earth mass case.

For a 10 Earth mass planet which is not embedded in the disk, the dynamical friction dominates the evolution of the semi-major axis up to about $50^\circ$ inclination. 
The evolution of the inclination is dominated by dynamical friction up to about $i\sim70^\circ$.
Beyond that, the aerodynamic drag becomes important. 

For a 100 Earth mass planet, dynamical friction dominates up to retrograde orbits with $i\sim130^\circ$. 
For orbits above $i \sim100^\circ$, the inclination damping time-scale is longer than the migration time-scale. 
This leads to interesting questions about the orbital evolution. 
From this analysis alone, we would expect an excess of planets with several tens of Earth masses on highly inclined orbits.

\section{Numerical simulations}\label{sec:numerical}
\begin{figure*}
\centering
\includegraphics[width=\textwidth]{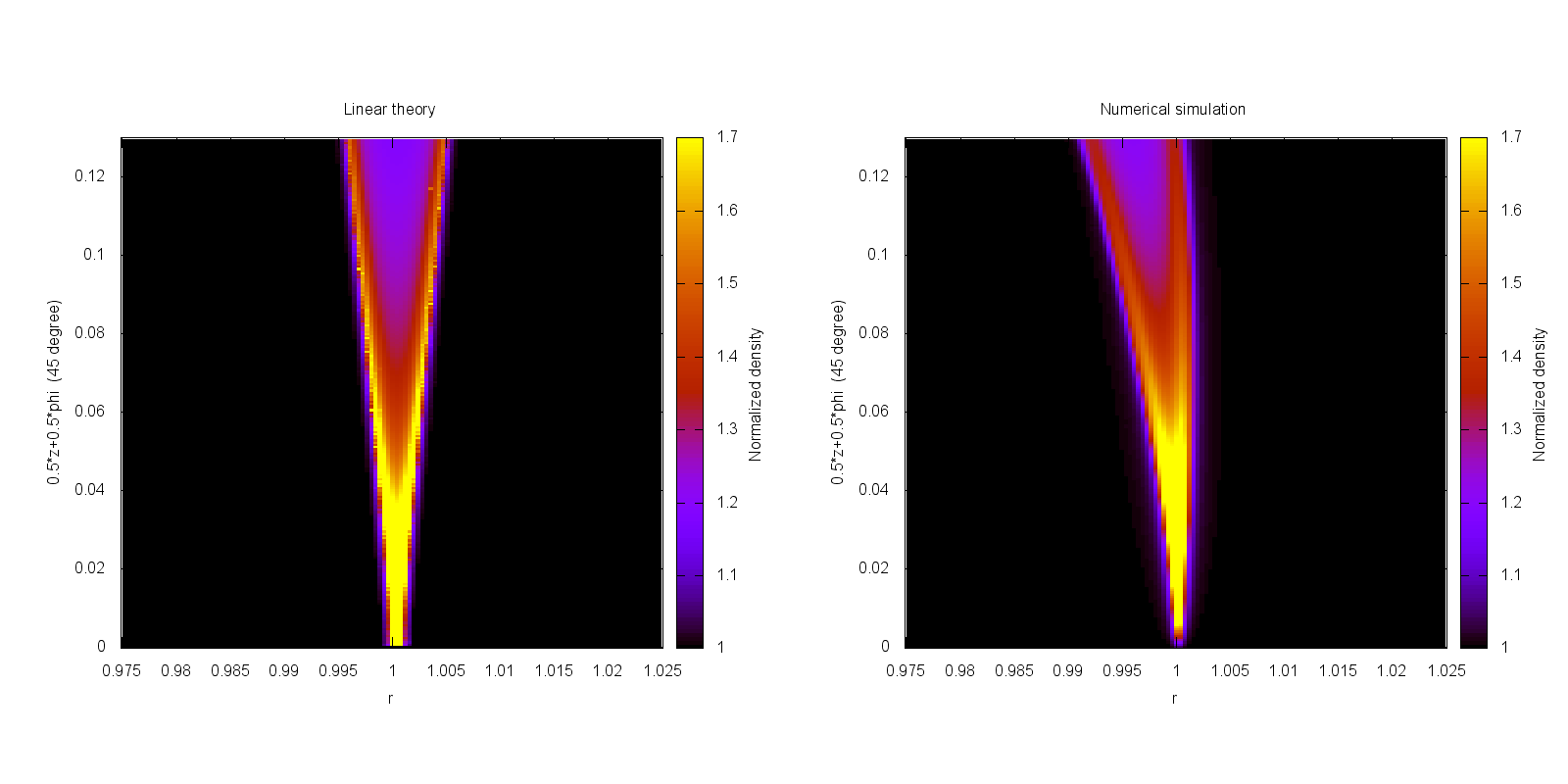}
\caption{Comparison of linear theory (left) and numerical simulations (right). Plotted are slices of the normalized density $\rho(\mathbf{x})/\rho_0(\mathbf{x}) = \alpha(\mathbf{x}) + 1$ for an Earth mass planet crossing the disk plane. The snapshot is taken when the planet crosses the mid-plane of the disk ($z=0$) and along the direction of the force (see Fig.~\ref{fig:setup}).\label{fig:comparison}}
\end{figure*}

\begin{figure}
\centering
\subfigure[Tangential force on the planet after removing the contribution due to the unperturbed disk as a function of time for three different inclinations. The mass of the planet is one Earth mass.]{\includegraphics[width=\columnwidth]{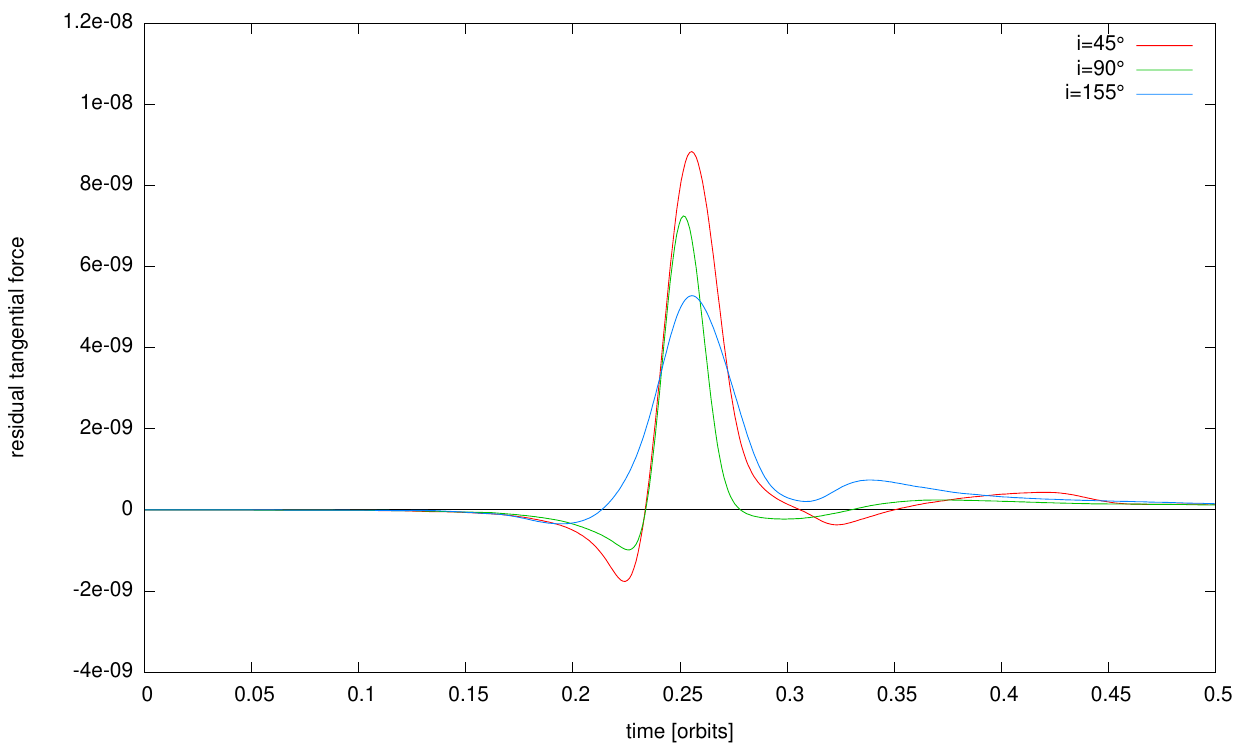}}
\subfigure[Fractional change in the planet's orbital energy after one crossing as a function of the inclination. The mass of the planet is one Earth mass.]{\includegraphics[width=\columnwidth]{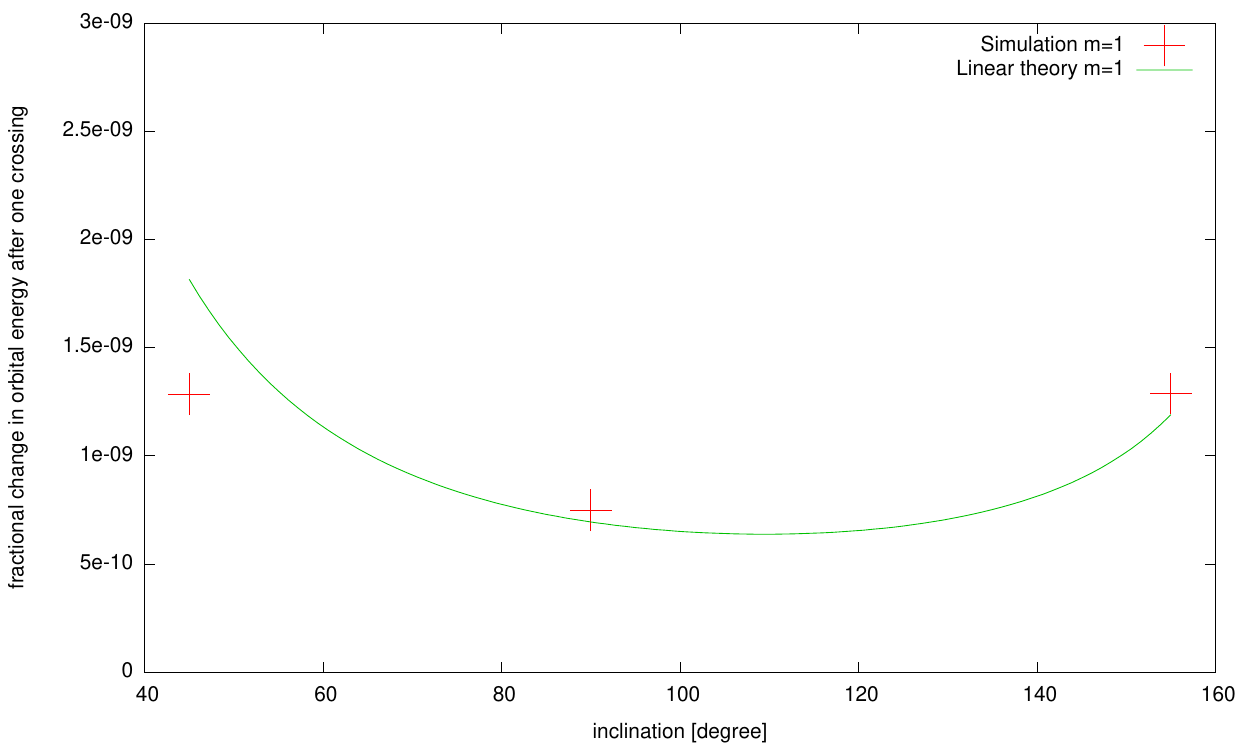}}
\subfigure[Fractional change in the planet's orbital energy after one crossing from numerical simulations and linear theory as a function of the softening length]{ \includegraphics[width=\columnwidth]{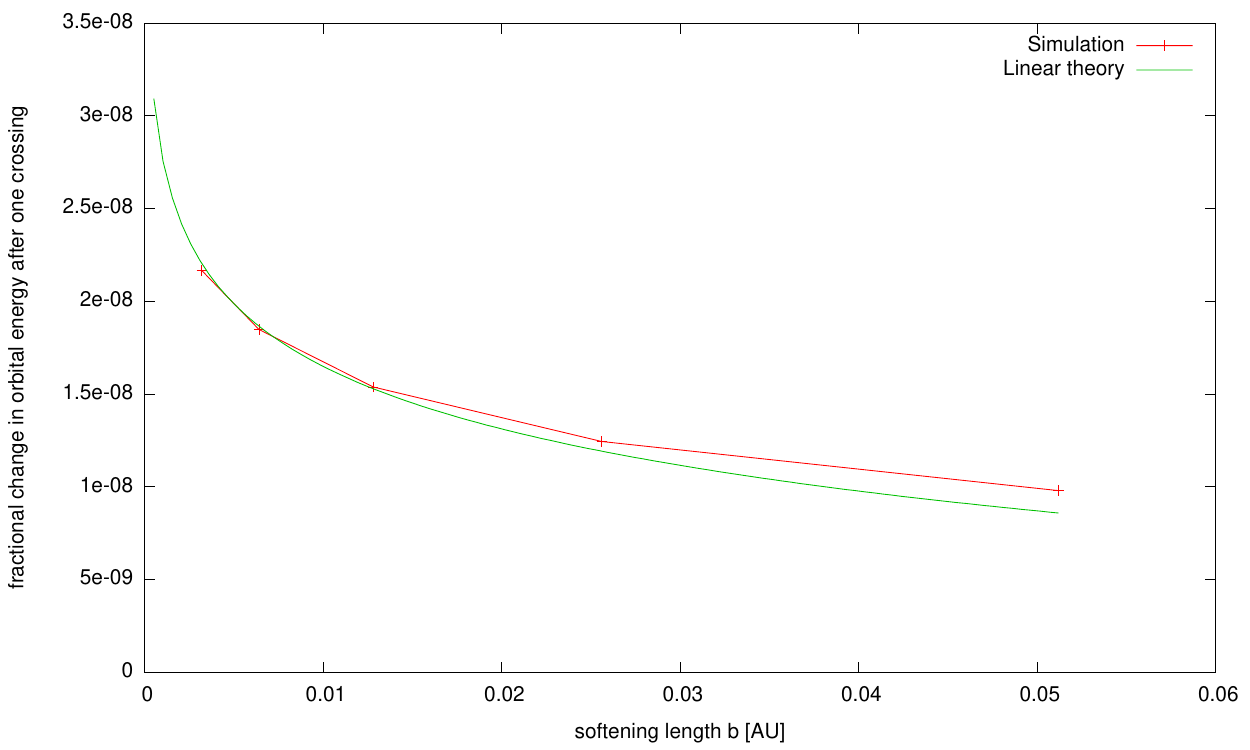} }
\caption{Results from the hydrodynamic simulations.\label{fig:three}}
\end{figure}
We ran high resolution, three dimensional hydro-dynamical simulations of a planet interacting with a proto-stellar disk to confirm the validity of the linear theory described above. 
In this paper we only simulate a small wedge of the disk to speed up calculations as we are only interested in one passage of the planet through the disk. 
The results presented in this section should be seen as a proof of concept, illustrating that it is possible to have a well defined and converged numerical simulation.
We will study interesting non-linear effects such as the question if the planet is able to open a gap and where the precise transition between the \cite{Tanakaetal2004} regime and dynamical friction is, in much more detail in an upcoming paper.

We use the Athena code in cylindrical coordinates with a resolution of up to $(N_r, N_\phi, N_z) = (768,1024,256)$. 
The size of the wedge is typically $(L_r, L_\phi, L_z) = (0.6,0.8,0.2)$. 
An isothermal equation of state is used with a sound-speed of $c_s=0.05$. 
The planet is on a $i=90^\circ$ orbit so that the Mach number is $M=28.3$.
Note that because of the CFL condition and $M\gg1$ the time-step in the simulation is very small, typically $\sim10^{-5}$.

Fig.~\ref{fig:comparison} shows a slice of the density perturbation $\rho(\mathbf{x})/\rho_0(\mathbf{x}) = \alpha(\mathbf{x}) + 1$ in the direction of the force ($45^\circ$ in the $z\phi$ plane for a planet on $i=90^\circ$). 
This is given analytically by Eq.~(\ref{eq:alphabar}) on the left and by the numerical simulation on the right. 
The overall agreement between the linear calculation and the numerical result is very good with two differences worth noting. 
First, the numerical solution is not symmetric with respect to $r=1$. 
This is due to the curvature of the orbit, whereas the analytic model on the left assumes that the orbit is on a straight line\footnote{Note that the plot is stretched in the $z\phi$ direction.}.
Second, the numerical solution is slightly smeared out compared to the analytic solution. 
This is due to numerical viscosity.

We show the outcome of simulations with a one Earth mass planet on an orbit with $i=45^\circ, 90^\circ, 155^\circ$ in Figs.~\ref{fig:three} (a) and (b).
Fig.~\ref{fig:three}~(a) shows the tangential force onto the planet after removing the part from the unperturbed disk as a function of time. 
The planet is crossing the mid-plane at $t=0.25$.
The differences in the force can be understood as follows.
The interaction of the planet with the disk in the $i=45^\circ$ case is long and strong (low $M$). 
For the $i=90^\circ$ case the interaction is shorter and weaker. 
For the $i=155^\circ$ case the interaction gets longer again (because the planet spends more time in the disk) but even weaker (high $M$). 

Integrating this component of the force along the trajectory of the planet gives the change in orbital energy, as calculated in Eq.~(\ref{eq:deltae}). 
We measure the change in specific orbital energy of the planet in each of those simulations.
The results are plotted in Fig.~\ref{fig:three}~(b), together with the theoretical predictions.
In the $i=45^\circ$ case there is a noticable difference which is most likely due to the curvature of the disk that becomes more important for small $i$. 

Finally, let us focus on the effect of the softening length.
The relevant scale in the problem is $H^2/a \sim0.0025 $. 
However, the expression for $\Lambda$ in Eq.~(\ref{eq:lambdashort}) suggests that there is only a weak dependence on the softening parameter $b$.
We therefore ran multiple simulations with different softening parameters to verify this. 
In Fig.~\ref{fig:three}~(c) we plot the fractional change in orbital energy of the planet after one disk crossing as a function of the softening parameter. 
The solid line is given by Eq.~(\ref{eq:deltae}). The agreement is very good, over more than one order of magnitude. 
Effects that haven't been considered in the linear calculation, such as shear and curvature of both the orbit and the disk, are all $O(H)\sim0.05$ or smaller, leading to the small error seen in Fig.~\ref{fig:three}~(c).

\section{Discussion}\label{sec:conclusions}
In this paper, we derived the linear equations for the evolution of highly inclined planets interacting with a proto-planetary disk.
Depending on the mass of the planet and the inclination, either dynamical friction or aerodynamic drag are dominant.
For certain inclinations, the migration time-scale is shorter than the inclination damping time-scale and even the disk lifetime. 

For very small inclinations ($i\lesssim5^\circ$), when the planet is embedded in the disk, other effects become important. 
The time-scales given by dynamical friction approach values comparable to standard type I migration as estimated by \cite{Tanakaetal2004}.
In reality there is a smooth transition between the two regimes (see Fig.~\ref{fig:tau}), although this is not captured by the calculation presented above.

Using a completely different approach, \cite{PapaloizouLarwood2000} calculate the damping timescales of highly eccentric planets. Their eccentricity timescale scales as $\sim e^3$. We have a scaling of $\sim i^4$ for small $i$ (see Eq.~\ref{eq:taui}). We can understand the difference by noting the interaction for an inclined planet can only take place while it is embedded in the disk, whereas a highly eccentric planet is always embedded in the disk.

A large fraction of high-mass, close-in planets (Hot Jupiters) has highly inclined orbits \citep{Triaud2010}.
On the other hand, low mass planet candidates in multiplanetary systems discovered by the Kepler mission seem to be aligned with low to moderate mutual inclinations \citep{Lissaueretal2011}.
If this trend is correct, then we can use the results from our calculation to predict that there are two distinct formation scenarios for low and high mass (and close-in) planets.
The timescale for realignment of small mass planets is very long. 
Therefore these objects have never been inclined otherwise we  would still see them on inclined orbits today.
Whatever process moved Hot Jupiters on inclined orbits does not operate on small mass planetary systems.

This calculation might also have interesting implications for other systems where dynamical friction occurs in the presence of a gaseous disk, such as for binary stars and super-massive black holes \citep[e.g.][]{Kim2010}. 

In a follow-up paper we will study the long-term and non-linear evolution with detailed global hydro-dynamical simulations. 
Using the results of this paper, a softening length of just slightly smaller than the disk scale height is sufficient to capture the relevant physical process.
Without these results, numerical simulations of planets on highly inclined orbits interacting with a proto-stellar disk would not be feasible.

\section*{Acknowledgments}
This manuscript has greatly benefited from the comments made by James Stone, Scott Tremaine, Woong-Tae Kim and an anonymous referee.
This work was supported in part by NSF grant AST-0807444 and by NASA grant NNX08AH83G.

\bibliographystyle{mn2e}
\bibliography{full}

\begin{thebibliography}{}

\bibitem[\protect\citeauthoryear{{Artymowicz}}{{Artymowicz}}{1994}]{Artymowicz%
1994}
{Artymowicz} P.,  1994, \apj, 423, 581

\bibitem[\protect\citeauthoryear{{Binney} \& {Tremaine}}{{Binney} \&
  {Tremaine}}{2008}]{BinneyTremaine2008}
{Binney} J.,  {Tremaine} S.,  2008, {Galactic Dynamics: Second Edition}.
Princeton University Press

\bibitem[\protect\citeauthoryear{{Bitsch} \& {Kley}}{{Bitsch} \&
  {Kley}}{2011}]{Bitsch2011}
{Bitsch} B.,  {Kley} W.,  2011, \aap, 530, A41+

\bibitem[\protect\citeauthoryear{{Burns}}{{Burns}}{1976}]{Burns1976}
{Burns} J.~A.,  1976, American Journal of Physics, 44, 944

\bibitem[\protect\citeauthoryear{{Chatterjee}, {Ford}, {Matsumura} \&
  {Rasio}}{{Chatterjee} et~al.}{2008}]{Chatterjee2008}
{Chatterjee} S.,  {Ford} E.~B.,  {Matsumura} S.,    {Rasio} F.~A.,  2008, \apj,
  686, 580

\bibitem[\protect\citeauthoryear{{Cresswell}, {Dirksen}, {Kley} \&
  {Nelson}}{{Cresswell} et~al.}{2007}]{Cresswell2007}
{Cresswell} P.,  {Dirksen} G.,  {Kley} W.,    {Nelson} R.~P.,  2007, \aap, 473,
  329

\bibitem[\protect\citeauthoryear{{Fabrycky} \& {Tremaine}}{{Fabrycky} \&
  {Tremaine}}{2007}]{FabryckyTremaine2007}
{Fabrycky} D.,  {Tremaine} S.,  2007, \apj, 669, 1298

\bibitem[\protect\citeauthoryear{{Kim} \& {Kim}}{{Kim} \&
  {Kim}}{2009}]{KimKim2009}
{Kim} H.,  {Kim} W.-T.,  2009, \apj, 703, 1278

\bibitem[\protect\citeauthoryear{{Kim}}{{Kim}}{2010}]{Kim2010}
{Kim} W.-T.,  2010, \apj, 725, 1069

\bibitem[\protect\citeauthoryear{{Lai}, {Foucart} \& {Lin}}{{Lai}
  et~al.}{2011}]{Lai2011}
{Lai} D.,  {Foucart} F.,    {Lin} D.~N.~C.,  2011, \mnras, 412, 2790

\bibitem[\protect\citeauthoryear{{Lissauer}, {Ragozzine}, {Fabrycky} \& {et
  al.}}{{Lissauer} et~al.}{2011}]{Lissaueretal2011}
{Lissauer} J.~J.,  {Ragozzine} D.,  {Fabrycky} D.~C.,    {et al.} 2011, \apjs,
  197, 8

\bibitem[\protect\citeauthoryear{{Marzari} \& {Nelson}}{{Marzari} \&
  {Nelson}}{2009}]{Marzari2009}
{Marzari} F.,  {Nelson} A.~F.,  2009, \apj, 705, 1575

\bibitem[\protect\citeauthoryear{{Ostriker}}{{Ostriker}}{1999}]{Ostriker1999}
{Ostriker} E.~C.,  1999, \apj, 513, 252

\bibitem[\protect\citeauthoryear{{Paardekooper}, {Baruteau}, {Crida} \&
  {Kley}}{{Paardekooper} et~al.}{2010}]{Paardekooper2010a}
{Paardekooper} S.-J.,  {Baruteau} C.,  {Crida} A.,    {Kley} W.,  2010, \mnras,
  401, 1950

\bibitem[\protect\citeauthoryear{{Paardekooper}, {Baruteau} \&
  {Kley}}{{Paardekooper} et~al.}{2011}]{Paardekooper2011b}
{Paardekooper} S.-J.,  {Baruteau} C.,    {Kley} W.,  2011, \mnras, 410, 293

\bibitem[\protect\citeauthoryear{{Papaloizou} \& {Larwood}}{{Papaloizou} \&
  {Larwood}}{2000}]{PapaloizouLarwood2000}
{Papaloizou} J.~C.~B.,  {Larwood} J.~D.,  2000, \mnras, 315, 823

\bibitem[\protect\citeauthoryear{{Simpson}, {Pollacco}, {Cameron}, {Hebrard},
  {Anderson} \& {Barros}}{{Simpson} et~al.}{2011}]{Simpson2011}
{Simpson} E.~K.,  {Pollacco} D.,  {Cameron} A.~C.,  {Hebrard} G.,  {Anderson}
  D.~R.,    {Barros} S.~C.~C. e.~a.,  2011, \mnras, pp 600--+

\bibitem[\protect\citeauthoryear{{Tanaka}, {Takeuchi} \& {Ward}}{{Tanaka}
  et~al.}{2002}]{Tanakaetal2002}
{Tanaka} H.,  {Takeuchi} T.,    {Ward} W.~R.,  2002, \apj, 565, 1257

\bibitem[\protect\citeauthoryear{{Tanaka} \& {Ward}}{{Tanaka} \&
  {Ward}}{2004}]{Tanakaetal2004}
{Tanaka} H.,  {Ward} W.~R.,  2004, \apj, 602, 388

\bibitem[\protect\citeauthoryear{{Triaud}, {Collier Cameron}, {Queloz},
  {Anderson}, {Gillon}, {Hebb}, {Hellier}, {Loeillet}, {Maxted}, {Mayor},
  {Pepe}, {Pollacco}, {S{\'e}gransan}, {Smalley}, {Udry}, {West} \&
  {Wheatley}}{{Triaud} et~al.}{2010}]{Triaud2010}
{Triaud} A.~H.~M.~J.,  {Collier Cameron} A.,  {Queloz} D.,  {Anderson} D.~R.,
  {Gillon} M.,  {Hebb} L.,  {Hellier} C.,  {Loeillet} B.,  {Maxted} P.~F.~L.,
  {Mayor} M.,  {Pepe} F.,  {Pollacco} D.,  {S{\'e}gransan} D.,  {Smalley} B.,
  {Udry} S.,  {West} R.~G.,    {Wheatley} P.~J.,  2010, \aap, 524, A25

\bibitem[\protect\citeauthoryear{{Weidenschilling}}{{Weidenschilling}}{1977}]{%
Weidenschilling1977}
{Weidenschilling} S.~J.,  1977, \apss, 51, 153

\end{thebibliography}
\label{lastpage}

\end{document}